\documentclass[pre,twocolumn,preprintnumbers,amsmath,amssymb,nofootinbib,floatfix,superscriptaddress]{revtex4}

\usepackage{graphicx,bm}

\makeatletter

\def\graphicscale{\twocolumn@sw{0.3}{0.4}}
\def\graphicthreescale{\twocolumn@sw{0.3}{0.4}}

\begin{document}

\title{Dynamic finite-size scaling after a quench at quantum transitions}

\author{Andrea Pelissetto}
\affiliation{Dipartimento di Fisica dell'Universit\`a di Roma
	``La Sapienza" and INFN, Sezione di Roma I, I-00185 Roma, Italy}

\author{Davide Rossini}
\affiliation{Dipartimento di Fisica dell'Universit\`a di Pisa
        and INFN, Largo Pontecorvo 3, I-56127 Pisa, Italy}

\author{Ettore Vicari} 
\altaffiliation{Authors are listed in alphabetic order.}
\affiliation{Dipartimento di Fisica dell'Universit\`a di Pisa
        and INFN, Largo Pontecorvo 3, I-56127 Pisa, Italy}

\date{\today}

\begin{abstract}
  We present a general dynamic finite-size scaling theory for the
  quantum dynamics after an abrupt quench, at both continuous and
  first-order quantum transitions.  For continuous transitions, the
  scaling laws are naturally ruled by the critical exponents and the
  renormalization-group dimension of the perturbation at the
  transition.  In the case of first-order transitions, it is possible
  to recover a universal scaling behavior, which is controlled by the
  size behavior of the energy gap between the lowest energy levels.
  We discuss these findings in the framework of the paradigmatic
  quantum Ising ring, and support the dynamic scaling laws by
  numerical evidence.
\end{abstract}

\maketitle


\section{Introduction}
\label{intro}

Understanding the quantum evolution of many-body systems is an
outstanding and intensely debated problem, starting from the dawn of
quantum mechanics.  Up to the beginning of the new millennium, issues
related to this topic were mostly considered as merely
academic~\cite{vonNeumann-29, Pauli-37, Niemeijer-67, Mazur-68,
  McCoy-70, McCoy-71}. However, the recent technological breakthroughs
in the realization, control, and readout of the coherent dynamics of
isolated quantum many-body systems for a significant amount of time
(as for ultracold atoms or trapped ions) have changed this point of
view~\cite{Bloch-08, GAN-14}.  As a matter of fact, they have
catalyzed a huge body of scientists, working both on the theoretical
and the experimental side.  In this context, the so-called {\it
  quantum quench} probably represents the simplest protocol in which a
system can be naturally put in out-of-equilibrium
conditions~\cite{Greiner-02, Weiss-06, Schmiedmayer-07, Trotzky-12,
  Cheneau-12, Schmiedmayer-12}

Several interesting issues have been deeply scrutinized in the recent
years for the quantum dynamics after a quench, stimulating a fervid
scientific activity.  They include the long-time relaxation and the
consequent spreading of correlations, the statistics of the work, the
mutual interplay of interactions and disorder, aging and coarsening
properties, short-time dynamic scaling, dynamical phase transitions,
to mention some of the most representative ones (see, e.g.,
Refs.~\cite{PSSV-11,NH-15,CTGM-15,CC-16,Heyl-17} and references
therein).  In this paper we focus on a further issue related to
quantum quench dynamics, that is the emergence of a dynamic
finite-size scaling (DFSS) in the quantum dynamics of an isolated
many-body system after a quench in proximity of a quantum phase
transition.  We put forward a DFSS theory in the appropriate limit,
which is valid at generic continuous quantum transitions (CQTs) and at
first-order quantum transitions (FOQTs).

Before entering the details of our investigation, let us formally
introduce the setting.  A quench protocol is generally performed
within a family of Hamiltonians, that are written as the sum of two
noncommuting terms:
\begin{equation}
  H(\lambda) = H_u + \lambda P .
  \label{hlam}
\end{equation}
The tunable parameter $\lambda$ enables us to modify the strength of
the {\em perturbation} $P$, e.g.,~a magnetic field term in a system of
interacting spins, with respect to the {\em unperturbed} Hamiltonian
$H_u$.  The idea of a quantum quench is to prepare the system in the
ground state of Hamiltonian~\eqref{hlam} associated with an initial
value $\lambda_0$, that is $|\Psi(0)\rangle \equiv
|0_{\lambda_0}\rangle$, and then suddenly change the parameter to
$\lambda\neq\lambda_0$. The resulting dynamic problem corresponds to
that of the quantum evolution driven by the Hamiltonian $H(\lambda)$,
starting from the particular initial condition of the ground state of
$H(\lambda_0)$, that is $|\Psi(t)\rangle = e^{-i H(\lambda)
  t}|0_{\lambda_0}\rangle$ (hereafter we will adopt units of $\hbar =
k_B = 1$).

We are interested in the quench dynamics occurring within the {\em
  critical} regime of a quantum transition. Thus, $H_u$ describes a
system at a CQT or a FOQT.  In the following we discuss the interplay
between the quench parameters $\lambda_0,\,\lambda$, and the finite
size $L$ of the system, assuming that both the initial ($\lambda_0$)
and final ($\lambda$) parameters keep the system close to the quantum
transition point. For this purpose, we define a DFSS limit as the
large-size and large-time limit, keeping the appropriate scaling
variables fixed. At CQTs such scaling variables are the combinations
$tL^{-z}$, $\lambda_0 L^{y_\lambda}$, and $\lambda L^{y_\lambda}$,
where $z$ and $y_\lambda$ are suitable critical exponents. Namely, $z$
is the dynamic exponent associated with the critical behavior of the
low-energy spectrum, and $y_\lambda$ is the renormalization-group (RG)
dimension of the parameter $\lambda$.  At FOQTs power laws may turn
into exponential laws related to the size dependence of the energy
gap.

The DFSS that we put forward is validated within the quantum Ising
chain, the paradigmatic model undergoing FOQTs and CQTs, when varying
its parameters.  In particular, we consider quench protocols
associated with variations of the longitudinal magnetic field coupled
to the order-parameter spin operators.  We present analytical and
numerical results for the off-equilibrium behavior of several
quantities, including the magnetization, the Loschmidt echo, which
measures the overlap between the evolved state and the initial state
of the system, and the bipartite entanglement entropy, quantifying the
quantum correlations between different spatial parts of the system.

The paper is organized as follows.  In Sec.~\ref{genscacqt} we put
forward the general DFSS theory for the quantum evolution after a
quench at CQTs.  In Sec.~\ref{isring} we specialize the discussion to
the quantum Ising ring: we introduce the model and we report the
expected scaling behavior of the magnetization and two-point
correlations.  We also thoroughly discuss the scaling of the Loschmidt
echo and of the bipartite entanglement entropy.  The predicted
asymptotic behaviors are then verified numerically.  In
Sec.~\ref{OFSSfo} we analytically derive the scaling functions for the
Ising ring along the FOQT line, by employing a two-level truncation of
the system's Hilbert space, and numerically show that they are
asymptotically exact, up to exponential corrections in the system
size.  Finally, Sec.~\ref{sec:concl} presents a summary, our
conclusions, and future perspectives.

\section{Dynamic scaling theory at a CQT}
\label{genscacqt}

We first recall that the theory of finite-size scaling (FSS) at
quantum transitions is well established, see, e.g.,~Ref.~\cite{CPV-14}
and references therein.  Briefly speaking, one can assume that a
$d$-dimensional quantum transition~\cite{Sachdev-book} is
characterized by two relevant parameters $\mu$ and $\lambda$, such
that they vanish at the critical point, with RG dimension $y_\mu$ and
$y_\lambda$, respectively.
The asymptotic FSS behavior of a generic observable $O$ with RG
dimension $y_o$ is thus given by
\begin{equation}
  O(L,\mu,\lambda) \approx 
  L^{-y_o} \, {\cal O}(\mu \, L^{y_\mu}, \lambda \, L^{y_\lambda}),
  \label{cqteq}
\end{equation}
where $L$ denotes the linear size of the $d$-dimensional system under
investigation.

In order to characterize the dynamic behavior after a quench,
we extend the FSS framework to the quench case. We consider a Hamiltonian
\begin{equation}
  H(\lambda) = H_c + \lambda P ,
  \label{hlambda}
\end{equation}
where $H_c$ is critical (for $\lambda=0$ the system undergoes a CQT)
and $\lambda$ is a control parameter associated with the relevant
perturbation $P$.  In the quench protocol we start from the ground
state of $H(\lambda_0)$. Then, at the reference time $t=t_0 = 0$, we
suddendly switch the coupling from $\lambda_0$ to $\lambda$ and follow
the subsequent evolution of the system.  In the following, we always
assume that $\lambda$ and $\lambda_0$ are sufficiently small, so that
the system is always close to the quantum transition point.

In order to write down the dynamic scaling ansatz for the post-quench 
behavior of the system, we introduce the scaling variables
\begin{equation}
\kappa(\lambda) = \lambda \, L^{y_\lambda},
\qquad  \theta = t \, L^{-z}, 
\label{kappathetact}
\end{equation}
where $t$ is the time, and $z$ is the dynamic exponent characterizing
the behavior of the energy differences of the lowest-energy states
and, in particular, the gap $\Delta\sim L^{-z}$.  A DFSS should emerge
in the infinite-volume limit $L\to\infty$, keeping $\theta$,
$\kappa_0\equiv\kappa(\lambda_0)$, and $\kappa \equiv \kappa(\lambda)$
fixed. Then, a generic global observable $O$, whose RG dimension at
the critical point is $y_o$, is expected to behave as
\begin{eqnarray}
O(t,L,\lambda_0,\lambda) &\approx& 
L^{-y_o} {\cal O}(\theta,\kappa_0,\kappa)\label{ocqt}\\
&=& L^{-y_o} {\cal O}(\theta,\kappa,\delta_\lambda),
\nonumber
\end{eqnarray}
where 
\begin{equation}
\delta_\lambda \equiv {\kappa\over \kappa_0} -1  
= {\lambda\over \lambda_0} -1 . 
\label{rla}
\end{equation}
An analogous scaling is expected for the correlation functions.  The
corrections to the above asymptotic DFSS laws are expected to decay as
negative powers of the size $L$. In the RG
language, they may, for example, arise from the presence of irrelevant
perturbations at the fixed point controlling the critical behavior.
Note that the equilibrium (ground-state) FSS behavior must be
recovered in the limit $\delta_\lambda\to 0$.  We also mention that a
similar dynamic scaling behavior was also proposed, and verified, in
the context of trapped bosonic gases (confined by a harmonic
potential) for quench protocols associated with the size of the
trap~\cite{CV-10}.

The Loschmidt amplitude quantifies the deviation of the post-quench
state at time $t > 0$ from the initial state before the quench. It is
defined as the overlap
\begin{equation}
A(t) = \langle 0_{\lambda_0} | \Psi(t) \rangle = 
\langle 0_{\lambda_0} | e^{-iH(\lambda) t} |0_{\lambda_0} \rangle.  
\label{ptdef}
\end{equation}
We introduce the rate function
\begin{equation}
 Q(t) = - \ln |A(t)|^2,
\label{qdef}
\end{equation}
which provides information on the so-called Loschmidt echo (in the
following we refer to $Q(t)$ as the Loschmidt echo).  Note that
$Q(t)=0$ implies the restoration of the initial quantum state.  We
conjecture that the time dependence of $Q(t)$ after the quench obeys
the DFSS behavior
\begin{equation}
  Q(t,L, \lambda_0, \lambda) \approx {\cal Q} (\theta, \kappa, \delta_\lambda).
\label{Lasca}
\end{equation}

We may also evaluate the work ${\cal L} = E - E_0$ necessary to
perform the instantaneous quench at $t=0$.
The energy $E$ injected into the system by
the quench is given by the expectation value of the post-quench
Hamiltonian $H(\lambda)$ on the initial (pre-quench) state
$|0_{\lambda_0}\rangle$:
\begin{eqnarray}
E &=& \langle 0_{\lambda_0} | H(\lambda) | 0_{\lambda_0} \rangle \nonumber\\
&=& \langle 0_{\lambda_0} | H(\lambda_0) | 0_{\lambda_0} \rangle + 
(\lambda-\lambda_0) \langle 0_{\lambda_0} | P | 0_{\lambda_0} \rangle .
\label{energy}
\end{eqnarray}
Since the initial energy is $E_0=\langle 0_{\lambda_0} | H(\lambda_0) |
0_{\lambda_0} \rangle$, we obtain
\begin{equation}
{\cal L} = E - E_0 = (\lambda-\lambda_0)  
\langle 0_{\lambda_0} | P | 0_{\lambda_0} \rangle. 
\label{calla1}
\end{equation}
In the DFSS limit, we can exploit the equilibrium FSS behavior, 
Eq.~\eqref{cqteq}, to evaluate the matrix element $\langle
0_{\lambda_0} | P | 0_{\lambda_0} \rangle$.  Assuming that $P =
\sum_{\bm x} P_{\bm x}$ is a sum of local terms, we have
\begin{equation}
{\cal L} \approx  (\lambda-\lambda_0) L^{d-y_p} \, f_p(\kappa_0) ,
\label{calla2}
\end{equation}
where $f_p$ is the equilibrium FSS function associated with the
observable $P/L^{d}$.  Taking into account the relation~\cite{Sachdev-book}
\begin{equation}
y_p + y_\lambda = d+z
\label{ypyl}
\end{equation}
between the RG dimensions of $\lambda$ and of the associated perturbation
$P$, the scaling behavior of the work ${\cal L}$ can be eventually
written as
\begin{equation}
{\cal L} \approx  L^{-z} \delta_\lambda  \, \kappa_0 \, f_p(\kappa_0) .
\label{calla3}
\end{equation}

We may also consider the large-volume limit of the above scaling
behaviors. If $O$ is an intensive variable that has a finite limit for
$L\to\infty$ at $\lambda,\lambda_0\not=0$, from Eq.~\eqref{ocqt} we
obtain the infinite-volume dynamic scaling behavior
\begin{equation}
O(t,L\to\infty,\lambda_0,\lambda) \approx \lambda^{-y_o/y_\lambda}
{\cal O}_\infty(\lambda^{z/y_\lambda} t,\delta_\lambda),
\label{ocqtli}
\end{equation}
which is valid for $\lambda,\lambda_0\to 0$, and $t\to\infty$,
keeping $\lambda^{z/y_\lambda} t$ and
$\delta_\lambda$ fixed. For $L\to\infty$, the 
work grows as the volume, which implies 
\begin{eqnarray}
&& f_p(\kappa_0) \sim \kappa_0^{y_p/y_\lambda}, \qquad 
   \kappa_0 \to \infty, \nonumber \\
&& {\cal L} \sim  L^{d} \delta_\lambda  \,\lambda_0^{1 + y_p/y_\lambda}.
\label{calla2li}
\end{eqnarray}

We finally remark that the above DFSS arguments can be
straightforwardly extended to more complicated quench protocols, for
example when they involve changes of both relevant parameters $\mu$
and $\lambda$.

\section{Scaling across the CQT of the Ising ring}
\label{isring}

To fix the ideas, we now demonstrate how a DFSS behavior emerges along
a sudden quench of the simplest paradigmatic quantum many-body system,
exhibiting a nontrivial zero-temperature behavior: the one-dimensional
quantum Ising chain in the presence of a transverse field.  Namely, we
show how to describe the interplay between the various parameters of
the quench protocol and the finite size of the system in an
appropriate DFSS limit.

\subsection{Hamiltonian model and quench protocol}
\label{Isingqupro}

The Hamiltonian of a quantum Ising ring is
\begin{equation}
H_{\rm Is} = - \sum_{x=1}^{L} \left[ J \,\sigma^{(3)}_x \sigma^{(3)}_{x+1} 
+ g\, \sigma^{(1)}_x\right],
\label{hedef}
\end{equation}
where, on each site $x$ of the chain, the spin variables ${\bm
  \sigma}\equiv (\sigma^{(1)},\sigma^{(2)},\sigma^{(3)})$ are the
Pauli matrices, and ${\bm \sigma}_{L+1}~=~{\bm \sigma}_{1}$ denotes
periodic boundary conditions.  The parameters $J$ and $g$ respectively
denote a ferromagnetic nearest-neighbor interaction (we assume $J=1$)
and the transverse field strength (we assume $g>0$).
A CQT occurs at $g=1$, separating a disordered ($g>1$) from an ordered
($g<1$) quantum phase~\cite{Sachdev-book}. This CQT belongs to the
two-dimensional Ising universality class with critical exponents
$\nu=1$, $\eta=1/4$, and $z=1$, which are associated with the
diverging length scale, the behavior of the two-point function at the
critical point, and the energy gap at the
transition, respectively.

In the following, we wish to analyze the quantum dynamics arising from
a quench protocol associated with an external magnetic field along the
longitudinal direction.  We thus add a magnetic perturbation $P =
-\sum_x \sigma^{(3)}_x$ to Eq.~\eqref{hedef}, that is we consider
\begin{equation}
  H(\lambda) = H_{\rm Is} - \lambda \sum_{x=1}^{L} \sigma_x^{(3)}.
\label{hla}
\end{equation}
In the quench protocol we start at $t=0$ from the ground state of the
system, at the parameter value $\lambda_0$, and suddenly change it to
$\lambda\neq \lambda_0$. Then we consider the time evolved state
$|\Psi(t)\rangle$.  The quantum evolution is characterized by the time
dependence of observables computed at $t>0$, such as the magnetization
$M$ and the connected correlation function
\begin{align}
M(t,L,\lambda_0,\lambda) = {1\over L} \langle\Psi(t)| \sum_{x=1}^L 
\sigma_x^{(3)}|\Psi(t) \rangle,
\label{mtdef}\\
G_c(x-y,t,L,\lambda_0,\lambda) = \langle \Psi(t) | \sigma^{(3)}_x 
\sigma^{(3)}_y | \Psi(t) \rangle_c,
\label{Gdef}
\end{align}
respectively.  We used the translation invariance to infer the space
dependence of $G_c$.

\subsection{Scaling behavior}
\label{qIstr}

We now focus on the dynamics arising from the quench protocol
when the unperturbed Hamiltonian $H_{\rm Is}$ is at the CQT point,
that is for $g=1$.  The parameters $\lambda_0$ and $\lambda$ are
assumed to be sufficiently small to maintain the system in the
critical region.

In the DFSS theory put forward in Sec.~\ref{genscacqt}, the
relevant scaling variables are reported in Eq.~\eqref{kappathetact}.
Specializing to the one-dimensional Ising model, we have 
\begin{equation}
y_\lambda =(d+z+2-\eta)/2 = 15/8,\qquad z=1.
\label{critexp}
\end{equation}
The magnetization~\eqref{mtdef} obeys the scaling
behavior
\begin{equation}
  M(t,L,\lambda_0,\lambda) \approx 
  L^{-\beta/\nu} {\cal M}(\theta,\kappa,\delta_\lambda),
  \label{mcheck}
\end{equation}
where $\beta$ denotes the magnetization critical exponent, and thus
$\beta/\nu=1/8$.  One might also consider other observables, such as
the connected two-point function~\eqref{Gdef}, that should scale as 
\begin{equation}
G_c(x,t,L, \lambda_0,\lambda) \approx 
L^{-\eta/\nu} {\cal G}(X,\theta,\kappa,\delta_\lambda),
\label{gdfss}
\end{equation}
where $\eta/\nu=1/4$, and $X\equiv x/L$. The corresponding length
scale, defined, for example, from the second moment of $G_c$, is
expected to behave as
\begin{equation}
\xi(t,L,\lambda_0,\lambda) \approx 
L \;\Xi(\theta,\kappa,\delta_\lambda).
\label{xifss}
\end{equation}
The work associated with the quench is expected to scale as in
Eq.~\eqref{calla3}. We note that, for a sudden change of the sign of
the magnetic field, i.e., for $\lambda = - \lambda_0$
(correspondingly, $\delta_\lambda=-2$), the ground-state energies of
the initial and final Hamiltonians are equal, due to the symmetry of
the Ising ring. Thus, in this case the work ${\cal L}$ also provides
the energy difference between the state of the system and that of the
ground state during the post-quench quantum evolution.

The asymptotic scaling behaviors are expected to be approached with
power-law suppressed corrections. In the quantum Ising ring without
boundaries (i.e., with periodic boundary conditions) scaling
corrections to the equilibrium FSS laws~\cite{CPV-14} usually decay
as $L^{-\omega}$, where $\omega = 2$ is the leading scaling-correction
exponent~\cite{CCCPV-00}.  As we shall observe in Sec.~\ref{numres},
corrections are compatible with an $L^{-2}$ behavior also in
out-of-equilibrium conditions ($\theta \neq 0$).  However, we cannot
exclude the appearance in the DFSS case of new types of scaling
corrections that decay with a smaller power of the lattice size,
originating, for instance, from the breaking of the time-translation
invariance due to the initial condition of the quench protocol.

Analogous DFSS laws can be written for quenches arising from the
sudden change of a local perturbation. For instance, one can replace
Eq.~\eqref{hla} with
\begin{equation}
  H(\lambda) = H_{\rm Is} - \lambda \sigma_1^{(3)} ,
\label{eq:Hlocal}
\end{equation}
where the perturbation is on a single site only.
In this case, the local magnetization on site $x$ should behave as
\begin{equation}
  M_x(t,L,\lambda_0,\lambda) \approx L^{-\beta/\nu}
  {\cal M}_l(x_p/L,\theta,\kappa,\delta_\lambda),
\label{mltts}
\end{equation}
where the RG dimension entering the definition of $\kappa$,
cf. Eq.~\eqref{kappathetact}, is
$y_\lambda=1/2$ (see Ref.~\cite{PRV-18} and references therein).
As a consequence, the DFSS behavior of its spatial average should be
\begin{equation}
M(t,L,\lambda_0,\lambda) \approx L^{-\beta/\nu} 
{\cal M}_a(\theta,\kappa,\delta_\lambda).
\label{mlttscon}
\end{equation}

Finally, we point out that the above DFSS arguments apply also to the
case in which the quench protocol is associated with a transverse
magnetic field, i.e.,~with the perturbation $P_t= \sum_{x=1}^{L}
\sigma_x^{(1)}$. In this case, the RG dimension of the perturbation is
simply $y_\lambda=1/\nu=1$.  For a transverse field the magnetization
vanishes by symmetry, but one may consider the two-point function and
the corresponding correlation length. We mention that some results for
quantum quenches involving the transverse field in the infinite-size
limit are reported in Refs.~\cite{RSMS-09,CEF-11,HPK-13}.

\subsection{Numerical results}
\label{numres}

\begin{figure}[!t]
\includegraphics[width=0.95\columnwidth]{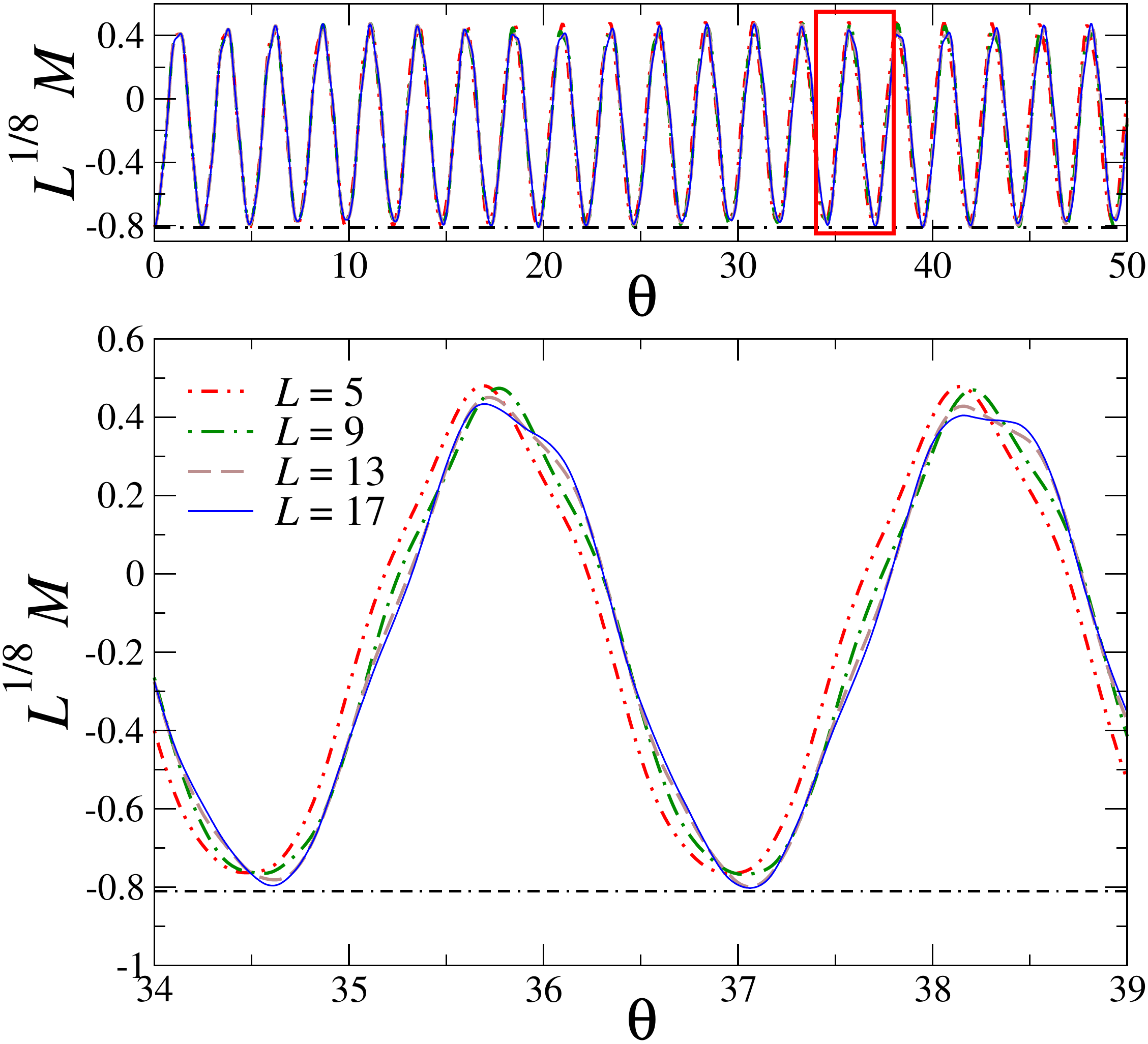}
\caption{Rescaled magnetization for the quantum critical Ising ring
  versus the rescaled time $\theta$, for $\kappa = 1$, $\delta_\lambda
  = -2$, and several values of $L$.  The bottom panel shows a
  magnification of the upper panel for $34 \leq \theta \leq 39$.  The
  horizontal dotted-dashed line indicates the infinite-size limit
  value of the static magnetization at $\kappa_0$.  Notice a clear
  trend towards an asymptotic oscillatory function with increasing
  $L$, thus confirming DFSS.}
\label{fig:CQT_Thetavar}
\end{figure}

In order to validate the DFSS predictions outlined above, we now
present the results of numerical simulations of the dynamics of the
Ising ring with Hamiltonian~\eqref{hla}, after a sudden quench in
$\lambda$.  For our purposes, it has been sufficient to consider
systems of moderate sizes (up to $L \approx 23$ sites). An exact
diagonalization approach has been used for systems with $L \leq 13$,
while Lanczos diagonalization followed by a fourth-order
Suzuki-Trotter decomposition of the unitary-evolution operator, with
time step $dt = 10^{-2}$, was employed for larger sizes ($14 \leq L
\leq 23$).

We start from the analysis of the magnetization defined in
Eq.~\eqref{mtdef}.  The numerical data of Fig.~\ref{fig:CQT_Thetavar},
corresponding to fixed values of $\kappa=1$ and $\delta_\lambda = -2$,
show that the product $L^{1/8}M$, as a function of the rescaled time $\theta$,
clearly approaches an asymptotic
function with increasing $L$.  This confirms the DFSS prediction~\eqref{mcheck}.
Convergence seems to be notably fast with $L$. An
oscillatory behavior, with the emergence of wiggles in proximity of
the peaks, clearly appears already for a moderately large size.
However, when zooming in the figure, a rather complicated pattern
emerges, signaling that the dynamics cannot be trivially obtained by
using an effective few-level description of the system.  As we shall
see later, it is however possible to extrapolate an asymptotic scaling
behavior at any value of $\theta$, which takes into account all these
features.

Notice also that the pseudo-sinusoidal trend persists at long times,
without appreciable damping in the oscillation amplitude.  Indeed, we
checked that the magnetization comes back periodically in time to a
value that is very close (although not equal) to the initial value,
whose extrapolated infinite-size limit is plotted as an horizontal
line in the figure.  The absence of a stationary large-$\theta$ limit
reflects the lack of thermalization, which is expected for this kind
of quench in the longitudinal field of the otherwise integrable Ising
ring.

\begin{figure}[!t]
\includegraphics[width=0.95\columnwidth]{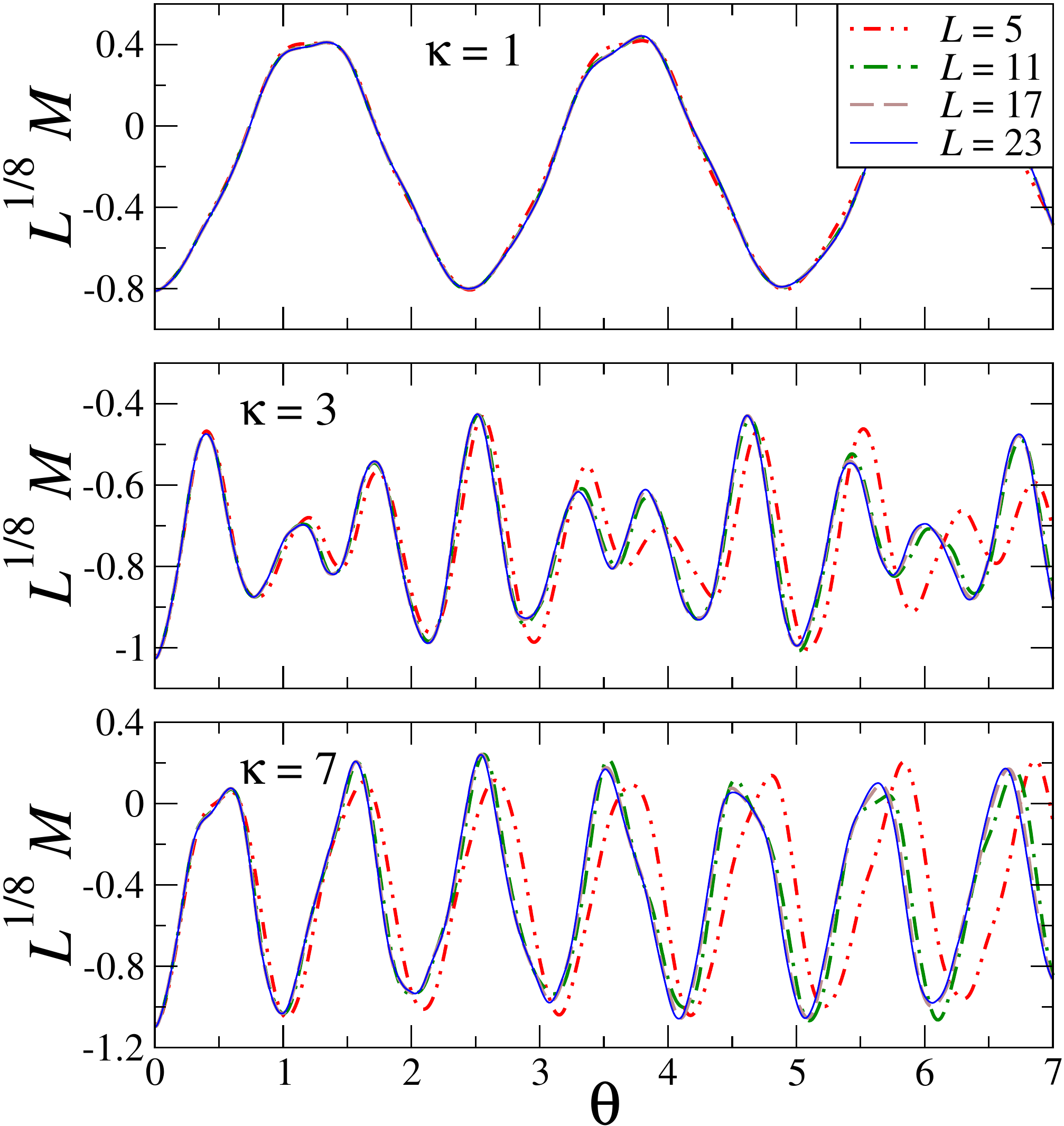}
\caption{Same plot as in Fig.~\ref{fig:CQT_Thetavar}, but for three
  different values of $\kappa$: $1$ (upper), $3$ (middle), $7$ (bottom
  panel).  In all cases the convergence to a scaling behavior, in the
  large-$L$ limit, is clearly visible.  Data in the upper panel are
  for the same parameters as in Fig.~\ref{fig:CQT_Thetavar}, but on a different
  time scale.}
\label{fig:Magnet_kvar}
\end{figure}

In Fig.~\ref{fig:Magnet_kvar} we change the value of $\kappa$, keeping
$\delta_\lambda=-2$.  Similar patterns emerge, all of them exhibiting
convergence to an asymptotic function, thus agreeing with the DFSS
prediction~\eqref{mcheck}.  It is tempting to compare the emerging
temporal features with those observed in Fig.~\ref{fig:CQT_Thetavar}:
for example, at $\kappa=3$, we observe a less regular pattern, with a
(pseudo-) periodicity which differs from the previous case.  For
$\kappa=7$ a more regular trend seems to reappear, although with a
much smaller period.  This behavior with $\kappa$ has to be ascribed
to the degree of commensurability of the injected energy with the
spectrum of the system.  Notice also that, for fixed $\theta$, the
approach to the asymptotic scaling behavior appears to be slower for
larger $\kappa$.

\begin{figure}[!t]
\includegraphics[width=0.9\columnwidth]{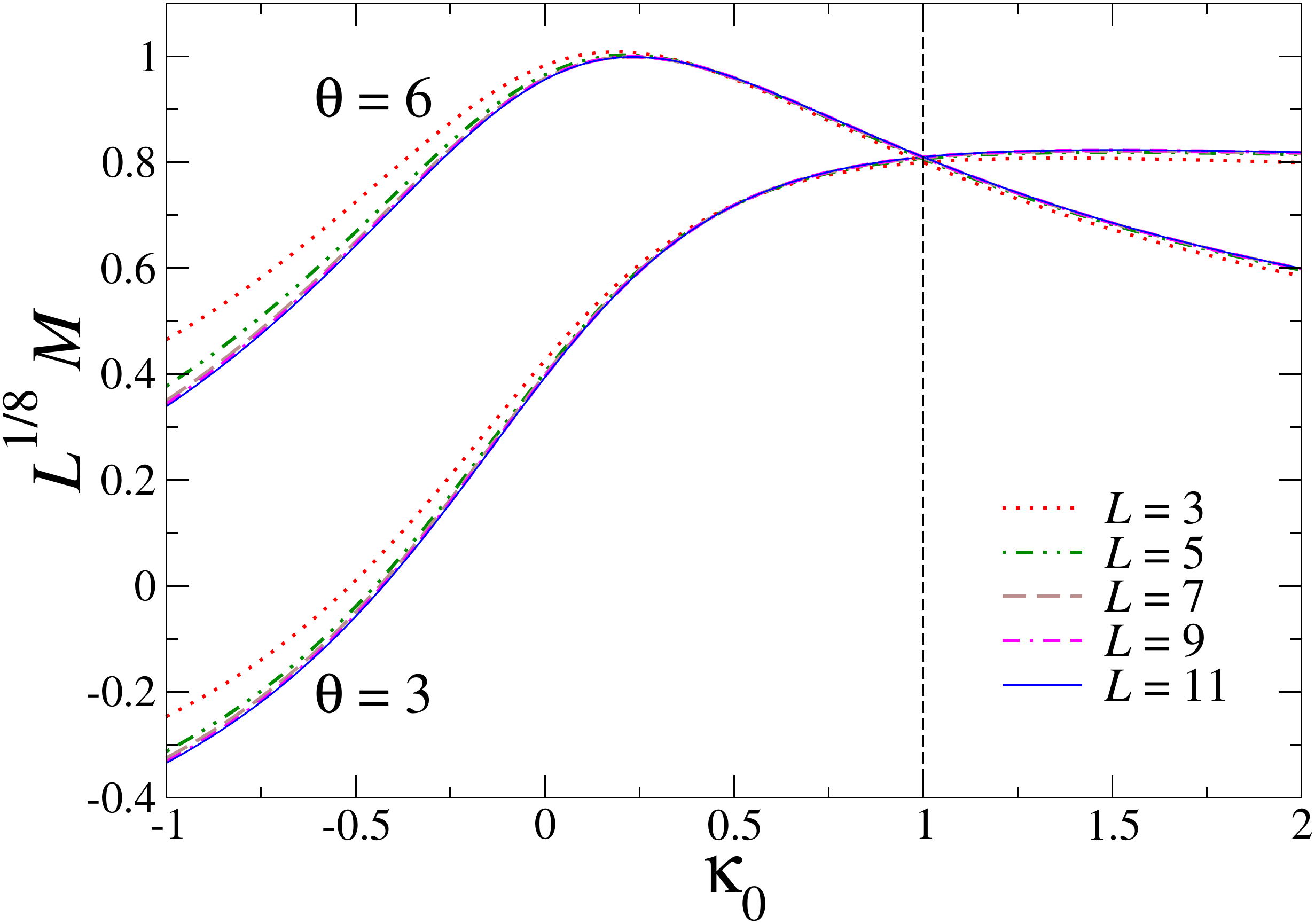}
\caption{Magnetization for fixed $\kappa=1$ and for two different
  rescaled times $\theta$.  The curves are plotted against the
  rescaled parameter $\kappa_0$, which is used to compute the initial
  state.  Notice that, at $\kappa_0=\kappa=1$, the equilibrium
  behavior is recovered (vertical dashed line).  As before, we observe
  that the curves at different size approach an asymptotic function,
  in accordance with our DFSS theory.}
\label{fig:CQT_Kivar}
\end{figure}

DFSS can also be checked as a function of the initial state, that is,
of the value of $\kappa_0$ before the quench. This is what we have
done in Fig.~\ref{fig:CQT_Kivar}, where we display the magnetization
after a quench at fixed rescaled time $\theta$ and $\kappa=1$, as a
function of $\kappa_0$ (thus, $\delta_\lambda = \kappa / \kappa_0-1 =
\lambda/\lambda_0 - 1$ is now changing).  Two values of $\theta$ are
shown. The various curves spotlight the emergence of a scaling
behavior, in a way similar to the previous cases as a function of
$\theta$. Obviously they intersect at the equilibrium point, which is
located at $\kappa_0=\kappa$, i.e., $\delta_\lambda=0$.

We have performed additional numerical simulations (not shown) for
several other choices of the scaling variables $\kappa_0, \,
\delta_\lambda, \, \theta$, confirming a similar fast convergence with
$L$ to the asymptotic functions, obeying the DFSS theory.

\begin{figure}[!t]
\includegraphics[width=0.95\columnwidth]{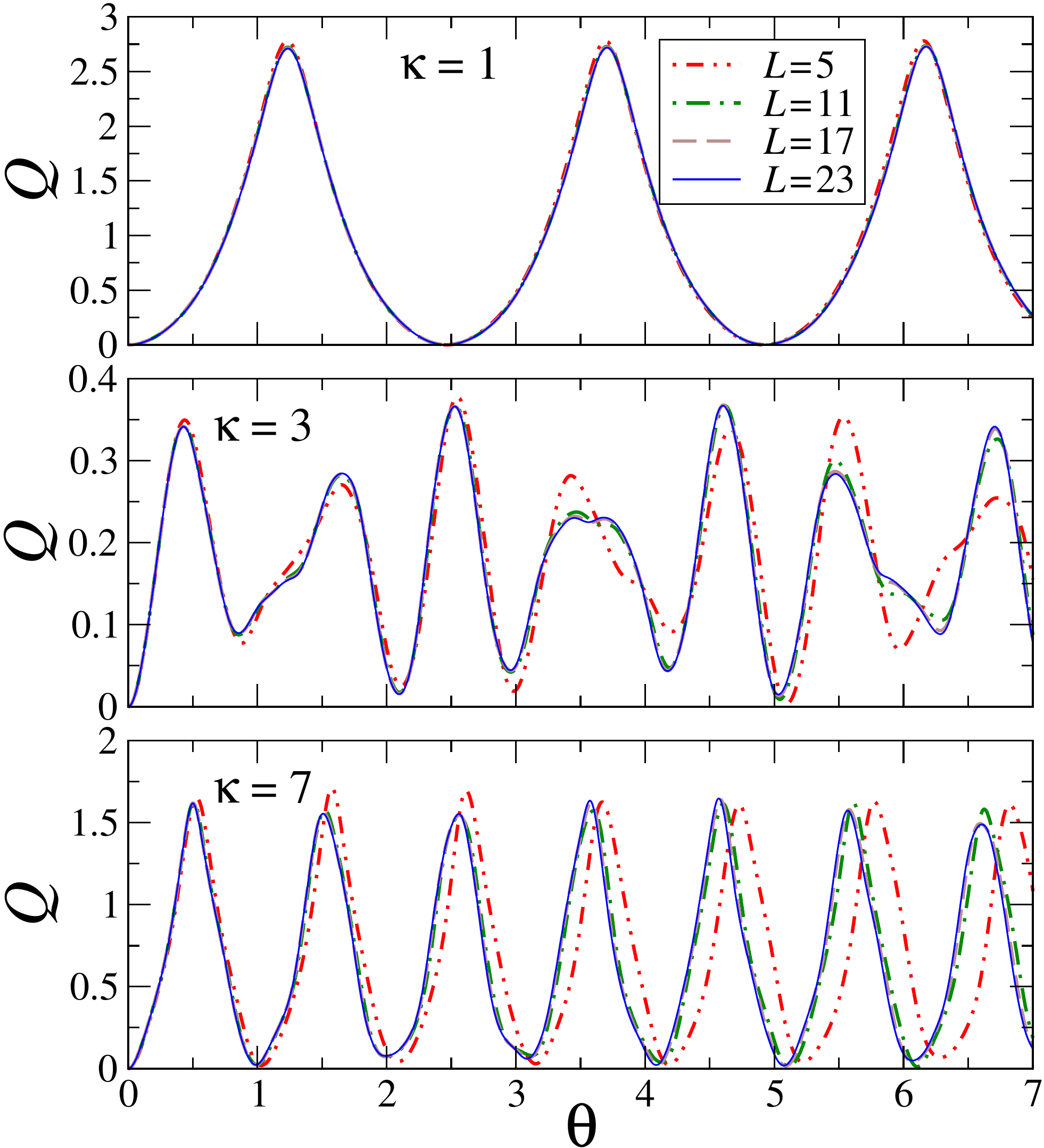}
\caption{Temporal behavior of the Loschmidt echo $Q(t)$ defined in 
  Eq.~\eqref{qdef} for $\delta_\lambda = -2$, and three
  different values of $\kappa$: $1$ (upper), $3$ (middle), $7$ (bottom
  panel), Convergence to a scaling function, in the large-$L$ limit,
  is clearly visible.}
\label{fig:LEcho_kvar}
\end{figure}

Let us now switch to the analysis of the Loschmidt echo $Q(t)$ defined
in Eq.~\eqref{qdef}.  Numerical data are plotted in
Fig.~\ref{fig:LEcho_kvar}; the emerging pattern is similar to that of
the magnetization, although quantitatively presenting different
features.  The data fully support the DFSS predicted by the scaling
equation~\eqref{Lasca}. We note in particular the evidence of
quasi-complete revivals of the quantum states along the quantum
evolution, when $Q(t)\ll 1$.

\begin{figure}[!t]
\includegraphics[width=0.98\columnwidth]{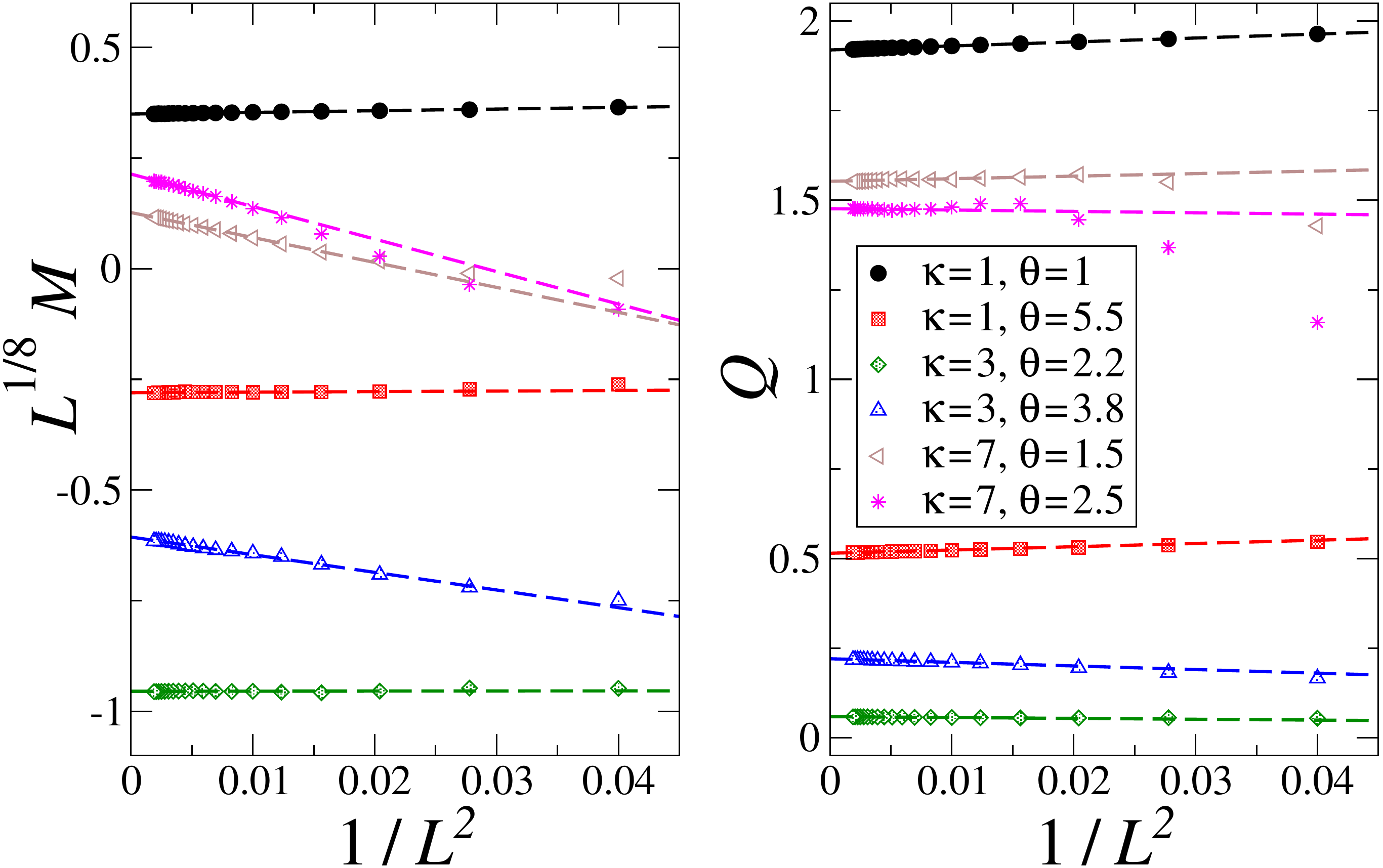}
\caption{Behavior of the rescaled magnetization $L^{1/8} M$ (left) and
  Loschmidt echo $Q(t)$ (right panel) with the system size, for
  different values of $\kappa$ and $\theta$, as indicated in the
  legend (data of Figs.~\ref{fig:Magnet_kvar} and~\ref{fig:LEcho_kvar}
  have been used).  Data are plotted against $1/L^2$: straight dashed
  lines denote $O(L^{-2})$ fits of the numerical values (symbols) for
  large $L$, and are plotted to guide the eye towards the extrapolated
  infinite-size limit.}
\label{fig:CQT_scal}
\end{figure}

To better check the convergence to the asymptotic scaling behavior in
the $L\to \infty$ limit, we have explicitly analyzed the dependence of
the various quantities with the size, keeping the scaling variables
fixed.  The corresponding data for the magnetization and the Loschmidt
echo, plotted as functions of $1/L^2$, are displayed in
Fig.~\ref{fig:CQT_scal}, where we highlight few representative values
of $\kappa$ and $\theta$.  Finite-size corrections appear to be
substantially consistent with an $L^{-2}$ behavior, which is the trend
expected for the homogeneous Ising ring at equilibrium~\cite{CCCPV-00}
(as a matter of fact, we explicitly checked the excellent quality of an
$L^{-2}$ fit at $\theta=0$).  Notice that, on the scale of the figure,
the dependence on $L$ is barely visible, except for $\kappa=7$, where
finite-size corrections are more evident.

Finally, we consider the time evolution of the entanglement entropy of
bipartitions of the system, which quantifies the amount of quantum
correlations that are present between the two parts of the chain.
These are operatively calculated by means of the following procedure:
we divide the chain into two connected parts of length $\ell_A$ and
$L-\ell_A$ (for the sake of clarity, we always take $\ell_A = L/2$),
and compute the so-called von Neumann (vN) entropy
\begin{equation}
  S(\ell_A,L)=S(L-\ell_A,L) = -{\rm Tr}\, \big[ \rho_A \ln \rho_A \big].
  \label{vNen}
\end{equation}
Here ${\rm Tr} [\, \cdot \,]$ denotes the trace operation, while
$\rho_A = {\rm Tr}_{L \setminus A} \big[ |\psi\rangle \langle \psi|
  \big]$ is the reduced density matrix of subsystem $A$, with
$|\psi\rangle$ being the quantum state of the global chain.  The
asymptotic large-$L$ behavior of the ground-state bipartite
entanglement entropy of the quantum Ising ring at the critical point
$g=1$ and $\lambda=0$ is known to be~\cite{HLW-94,CC-04,JK-04}:
\begin{equation}
  S_c(\ell_A,L) = {1\over 6} \big[\ln L + \ln{\rm sin}(\pi\ell_A/L) +
    e \big] + O(L^{-2}),
\label{ccfo}
\end{equation}
where $e$ is a known constant.  Definition~\eqref{vNen} applies also
to the time-dependent case, allowing us to compute the bipartite vN
entropy $S(\ell_A,L,t,\lambda_0,\lambda)$ on the state
$|\Psi(t)\rangle$ resulting after the quench at $t=0$.
We consider, in particular, the case of a balanced bipartition, i.e.,
$\ell_A/L=1/2$.  Extending equilibrium scaling arguments, see,
e.g.,~Ref.~\cite{CPV-14}, we conjecture the DFSS behavior
\begin{eqnarray}
\Delta S_{1/2} &\equiv&
S(L/2,L,t,\lambda_0,\lambda) - S_{c}(L/2,L) 
\nonumber\\
&\approx& {\cal S}(\theta,\kappa,\delta_\lambda).
\label{esca}
\end{eqnarray}

\begin{figure}[!t]
\includegraphics[width=0.95\columnwidth]{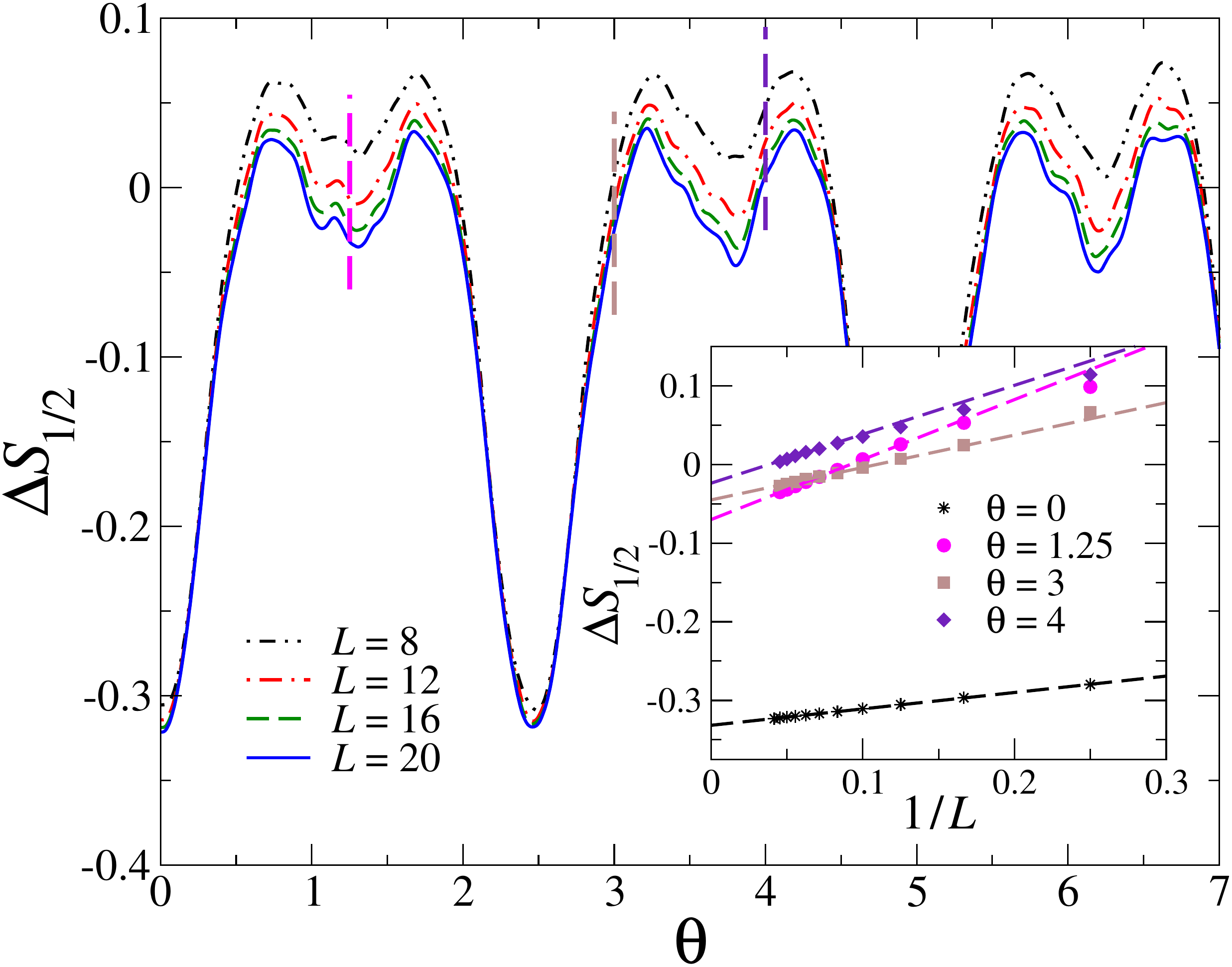}
\caption{Temporal behavior of the entanglement entropy for a balanced
  bipartition of $L/2$ sites, after a quench of the longitudinal field
  in the quantum Ising ring.  The inset displays the convergence with
  $L$ of the various curves (up to $L=22$), for three values of
  $\theta \neq 0$ (see the long-dash lines in the main panel), plotted
  against $1/L$. Additionally, black stars denote data corresponding to the
  equilibrium condition $\theta = 0$ (up to $L=24$).
  Dashed lines are numerical fits of the data at the largest available $L$.}
\label{fig:VNentro}
\end{figure}

DFSS is nicely supported by the time dependence of the half-chain vN
entropy, as shown in Fig.~\ref{fig:VNentro}.  We have also studied the
rate of approach to the asymptotic regime.  As spotlighted in the
inset of Fig.~\ref{fig:VNentro}, corrections to the asymptotic DFSS
behavior~\eqref{esca} generally scale as $1/L$.  This is true both for
the ground state of the initial Hamiltonian (see the black stars in
the inset, corresponding to $\theta=0$) and for the evolved state (we
report results for three different values of $\theta \neq 0$). These
corrections, and in particular those at $\theta=0$ corresponding to
the initial equilibrium ground state, are related to the so-called
conical corrections~\cite{CCP-10}. They are expected to be generally
$O(1/L)$ for the bipartite vN entanglement entropy around the CQT of
the quantum Ising chain, see e.g.~Ref.~\cite{CPV-14} for a detailed
discussion.  Note however that finite-size corrections for the
ground-state vN entropy decay as $1/L^2$ at the CQT point ($g=1$ and
$\lambda = 0$) for periodic boundary conditions, cf. Eq.~\eqref{ccfo},
where the leading conical correction cancels out.  Our numerical data
show that this cancellation does not occur for $\lambda\not=0$.

\section{Dynamic finite-size scaling along the FOQT line}
\label{OFSSfo}

In this section we extend the DFSS theory to FOQTs. Although the
presentation refers to the Ising ring with Hamiltonians~\eqref{hla}
or~\eqref{eq:Hlocal}, the results apply quite straightforwardly to
generic transitions.

For any $g<1$ (we assume $g>0$) the ground state of the Ising
Hamiltonian~\eqref{hedef} is doubly degenerate. The degeneracy is
lifted by the introduction of a longitudinal field, such as that
appearing in Eqs.~\eqref{hla}) and~\eqref{eq:Hlocal}. Therefore,
$\lambda = 0$ is a FOQT point, where the longitudinal magnetization $M
= L^{-1} \sum_{x=1}^{L} M_x$, with $M_x\equiv \langle \sigma_x^{(3)}
\rangle$, becomes discontinuous in the infinite-volume limit.  The
FOQT separates two different phases characterized by opposite values
of the (spontaneous) magnetization $m_0$ given by~\cite{Pfeuty-70}
\begin{equation}
\lim_{\lambda \to 0^\pm} \lim_{L\to\infty} M 
= \pm m_0, \qquad m_0 = (1 - g^2)^{1/8}.
\label{m0}
\end{equation}
In a finite system of size $L$, the two lowest states are superpositions
of two magnetized states $| + \rangle$ and $| - \rangle$ such that
\begin{equation}
\langle \pm | \sigma_x^{(3)} | \pm \rangle = \pm \,m_0 
\label{states}
\end{equation}
for all $x$.  Due to tunneling effects, the energy gap $\Delta$
vanishes exponentially as $L$ increases~\cite{Pfeuty-70,CJ-87}:
\begin{eqnarray}
\Delta(L) \approx 2 \left({1-g^2\over \pi L}\right)^{1/2} g^L ,
\label{deltalas}
\end{eqnarray}
while the differences $\Delta_i\equiv E_i-E_0$ for the higher excited
states $(i>1$) are finite for $L\to \infty$.

We consider a quench protocol in which $\lambda$ is suddenly varied.
To define the general DFSS laws, we proceed as in
Sec.~\ref{genscacqt}. First, we identify the relevant scaling
variables.  Arguments analogous to those reported in
Ref.~\cite{PRV-18} lead us to introduce the following quantities:
\begin{equation}
\kappa(\lambda) = {2 m_0 \lambda L^b \over \Delta(L)},\qquad
\theta = t \, \Delta(L),
\label{scavarfoqt}
\end{equation}
where $b=1$ for the homogenous perturbation $P$ of Eq.~\eqref{hla},
and $b=0$ for the local perturbation $P_l=-\sigma_1^{(3)}$ appearing
in Eq.~\eqref{eq:Hlocal}.  In particular, $\kappa(\lambda)$ is the
ratio between the energy associated with the longitudinal-field
perturbation, which is approximately $2m_0 \lambda L^b$, and the
energy difference $\Delta(L)$ of the two lowest states at $\lambda=0$.
Then, we may put forward the following DFSS for the magnetization:
\begin{equation}
M(t,L,\lambda_0,\lambda, L) =  
m_0 \, {\cal M}_{fo} (\theta,\kappa_0,\kappa) ,
\label{mcheckfoqt}
\end{equation}
where $\kappa_0\equiv\kappa(\lambda_0)$, $\kappa \equiv
\kappa(\lambda)$, and $m_0$ is given by Eq.~\eqref{m0}.  DFSS is
expected to hold for any $g<1$. In particular, the scaling function
${\cal M}_{fo} (\theta,\kappa_0,\kappa)$ is expected to be independent
of $g$, apart from trivial normalizations of the arguments.

The previous scaling relations can be straightforwardly extended to
any FOQT, by identifying the scaling variable $\kappa(\lambda)$ as the
ratio $\lambda E_p(L)/\Delta(L)$, where $E_p(L)$ is the energy
associated with the perturbation $P$ and $\Delta(L)$ is the energy
difference between the two lowest states at the transition point. The
second scaling variable $\theta$ is always defined as in
Eq.~\eqref{scavarfoqt}.

In the case of the quantum Ising ring, some scaling functions can be
exactly computed, exploiting a two-level truncation of the
spectrum~\cite{CNPV-14,PRV-18}.  As shown in Ref.~\cite{PRV-18}, in
the long-time limit and for large systems, the scaling properties in a
small interval around $\lambda=0$ (more precisely, for $m_0
|\lambda|\ll \Delta_2$) are captured by a two-level truncation, which
only takes into account the two nearly-degenerate lowest-energy
states.  The effective evolution is determined by the Schr\"odinger
equation~\cite{PRV-18}
\begin{equation}
i {d\over dt} \Psi(t) = H_{2}(\lambda) \Psi(t) ,
\label{sceq}
\end{equation}
where $\Psi(t)$ is a two-component wave function, whose components
correspond to the states $|+ \rangle$ and $|-\rangle$, and
\begin{eqnarray}
H_{2}(\lambda) = - \beta \, \sigma^{(3)} + \delta \, \sigma^{(1)}\,
. \label{hrtds}
\end{eqnarray}
Here $\beta = m_0 \lambda L^b$ and $\delta={\Delta/2}$, such that
$\kappa(\lambda)= 2\beta/\Delta$ and $\theta = 2 t \delta$. The
initial condition is given by the ground state of $H_2(\lambda_0)$,
i.e., by
\begin{equation}
\Psi(t=0) = \sin(\alpha_0/2) \, |-\rangle +  
\cos(\alpha_0/2) \, |+\rangle, \label{eigstatela0}
\end{equation}
with $\tan \alpha_0 = \kappa_0^{-1}$.  The quantum evolution can be
easily obtained by diagonalizing $H_{2}(\lambda)$, obtaining the
eigenstates
\begin{eqnarray}
&& |0\rangle = \sin(\alpha/2) \, |-\rangle +  
\cos(\alpha/2) \, |+\rangle, \label{eigstate0la}\\
&& |1\rangle =  \cos(\alpha/2) |-\rangle -
\sin(\alpha/2) \, |+\rangle, \label{eigstate1la}
\end{eqnarray}
where $\tan \alpha = \kappa^{-1}$, and the eigenvalue difference
\begin{equation}
\Delta_\kappa \equiv E_1 - E_0 = \Delta \; \sqrt{1 + \kappa^2}.
\end{equation}
Then, apart from an irrelevant phase, the time-dependent state evolves as
\begin{equation}
|\Psi(t)\rangle = 
\cos\left({\alpha_0-\alpha\over 2}\right) |0\rangle + 
e^{-i \Delta_\kappa t} \sin\left({\alpha_0-\alpha\over 2}\right) |1\rangle.
\label{psitfo}
\end{equation}
The magnetization is obtained by computing the
expectation value $\langle \Psi(t) | \sigma^{(3)} |\Psi(t) \rangle$.
It gives for the dynamic scaling function defined in 
Eq.~\eqref{mcheckfoqt}:
\begin{align}
  {\cal M}_{fo}(\theta,\kappa_0,\kappa) & =  \cos (\alpha-\alpha_0) \cos \alpha \nonumber \\
  & + \cos \big(\theta \sqrt{1+\kappa^2} \big) \sin(\alpha-\alpha_0) \sin \alpha\,.
  \label{m2lsca}
\end{align}
The approach to the asymptotic result is expected
to be exponential in the size of the system.

\begin{figure}[!t]
\includegraphics[width=0.95\columnwidth]{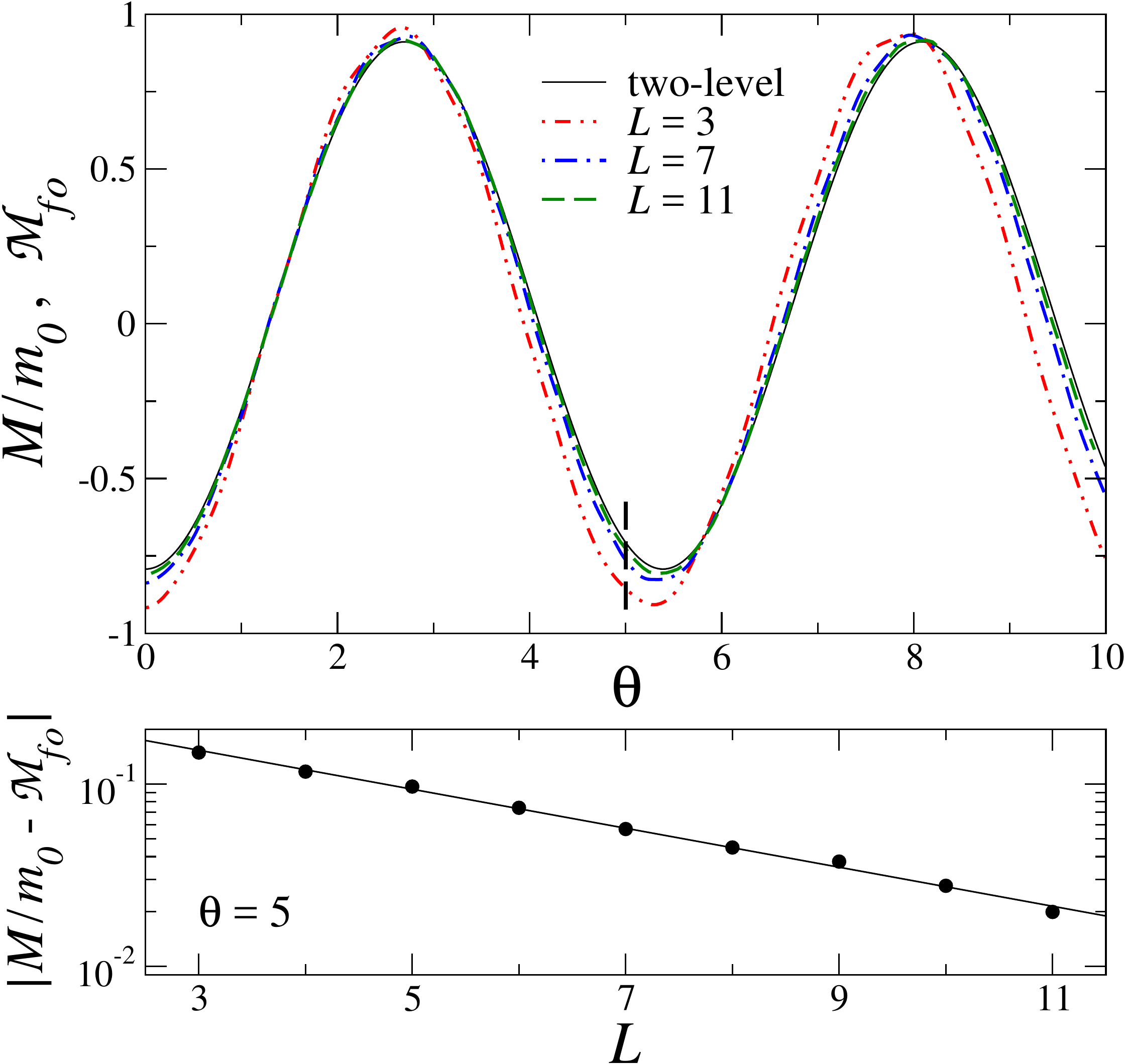}
\caption{Upper panel: Plot of the ratio $M/m_0$ for various system
  sizes (color dashed curves) and of the corresponding scaling function
  ${\cal M}_{fo}(\theta,\kappa_0,\kappa)$, reported in Eq.~\eqref{m2lsca}
  (black dashed curve), as functions of the rescaled time $\theta$,
  for fixed $\kappa_0 = -1.3, \, \kappa= 0.6$.
  Results are for the Ising ring at $g=0.9$.  Lower panel:
  difference between the numerically computed $M/m_0$ and the function
  ${\cal M}_{fo}$, as a function of $L$ at fixed $\theta = 5$.
  The straight line corresponds to an exponential fit of the data.}
\label{fig:FOQT}
\end{figure}

Figure~\ref{fig:FOQT} shows the behavior of the function ${\cal
  M}_{fo}(\theta,\kappa_0,\kappa)$, for fixed $\kappa_0, \kappa$ and
varying $\theta$, which displays the characteristic Rabi oscillations
naturally emerging in the dynamics of a two-level system.
Prediction~\eqref{m2lsca} is also compared with the estimates of the
magnetization obtained in numerical simulations.  Numerical data are
very close to ${\cal M}_{fo}$ already for small chain lengths, even if
data are obtained at $g=0.9$, thus relatively close to the CQT.  More
precisely, as shown in the lower panel, scaling corrections to the
two-level scaling prediction are exponentially suppressed with $L$.
The nice agreement confirms that, in the DFSS limit, the dynamics can be
faithfully approximated by truncating the Hilbert space to the two
lowest-energy states. If $\lambda$ and $\lambda_0$ have opposite sign
(a condition that is not necessary to observe DFSS), the two states
essentially correspond to the ground states of the initial and the
final Hamiltonian $H(\lambda_0)$ and $H(\lambda)$, respectively.

\section{Summary and Conclusions}
\label{sec:concl}

We have considered the dynamics of a quantum system subject to a
sudden change of a Hamiltonian parameter. Close to a quantum
transition, a DFSS behavior emerges from the interplay of the
parameters involved in the quench protocol and the size of the system.
In particular, we have considered a generic Hamiltonian $H(\lambda) =
H_c + \lambda P$, with $[H_c,P]\neq 0$, and focused on a sudden change
of the parameter $\lambda$ assuming that the pre- and post-quench
Hamiltonians remain in the {\em critical} regime of a quantum
transition. The DFSS limit is defined as the large-size and large-time
limit, keeping appropriate scaling variables fixed. At CQTs the scaling
variables are the combinations $tL^{-z}$, $\kappa_0 = \lambda_0
L^{y_\lambda}$ and $\kappa = \lambda L^{y_\lambda}$, where $z$ and
$y_\lambda$ are appropriate critical exponents, i.e.,~$z$ is the
dynamic exponent characterizing the size behavior of the energy gap,
and $y_\lambda$ is the RG dimension of the parameter $\lambda$. Note
that, for relevant perturbations for which $y_\lambda > 0$, the
parameters $\lambda_0$ and $\lambda$ have both a zero limit in the
scaling regime, thereby guaranteeing that the system is always in the
critical regime. It is also possible to include the effect of a small
finite temperature, assuming a Gibbs ensemble as initial condition, by
adding the scaling variable $\rho = TL^z$ as an additional argument of
the DFSS functions. The general theory applies also to FOQTs with the
only change that $\kappa$ should be defined as the ratio $\lambda
E_p(L)/\Delta(L)$, where $E_p(L)$ is the energy associated with the
perturbation $P$ and $\Delta(L)$ is the energy difference between the
two lowest-energy states.  In this case, it is possible for
$\kappa(L)$ to depend exponentially on $L$ as a consequence of the
finite-size behavior of the energy gap. We stress that the scaling
arguments we have presented are quite general.  Thus, they are expected
to apply to generic CQTs and FOQTs in any spatial dimension.

We have verified the DFSS theory in the quantum Ising chain, the
paradigmatic model undergoing FOQTs and CQTs, when varying its
parameters.  In particular, we have considered quench protocols
associated with changes of a longitudinal magnetic field coupled to
the order-parameter spin operator.  We have presented analytical and
numerical results for the off-equilibrium behavior of several
quantities, including the magnetization, the Loschmidt echo, and the
bipartite entanglement entropy.  The results fully support the
predictions of the DFSS theory we put forward.

A related important issue regards thermalization, that is, whether the
system has a local thermal-like behavior at an asymptotically long
time after the quench.  Understanding under which circumstances this
occurs is a highly debated issue~\cite{NH-15}, which lies outside the
purpose of our analysis, being related to the integrability properties
of the Hamiltonian $H_c$, the mutual interplay of interactions and
inhomogeneities, and the nature of the spectrum. Naive scaling
arguments suggest that, if the quantum evolution leads to an effective
thermalization, the eventual effective temperature scales as $T
\approx L^{-z} f_T(\kappa,\delta_\lambda)$. More likely, an effective
thermalization may emerge in the large-volume limit (of nonintegrable
systems), keeping the parameters $\lambda_0,\,\lambda$ fixed, i.e.~in
the limit $\kappa\to\infty$, when the energy provided to the system
grows as the volume, as argued at the end of Sec.~\ref{genscacqt}.

Finally we comment on the fact that, as foreseen by the outcomes of
our numerical simulations, it is likely that the general DFSS theory
following a quantum quench, described in Sec.~\ref{genscacqt}, can be
verified even with systems of relatively small size (i.e.~of the order
of 10 spins).  Therefore, given the need for high accuracies without
necessarily reaching scalability to large sizes, we believe that the
available technology for probing the coherent quantum dynamics of
interacting systems, such as with ultracold atoms in optical
lattices~\cite{Greiner_2011}, trapped ions~\cite{Monroe_2010,
  MonroeB_2010, Monroe_2014, Roos_2014}, as well as Rydberg atoms in
arrays of optical microtraps~\cite{Rydberg_2016}, could offer a unique
playground where this theory can be reliably tested.


\begin{thebibliography}{99}

\bibitem{vonNeumann-29}
  J. von Neumann,
  Beweis des Ergodensatzes und des H-Theorems in der neuen Mechanik,
  Z. Phys. {\bf 57}, 30 (1929).

\bibitem{Pauli-37}
  W. Pauli and M. Fierz,
  \"Uber das H-Theorem in der Quantenmechanik,
  Z. Phys. {\bf 106}, 572 (1937).

\bibitem{Niemeijer-67}
  T. Niemeijer,
  Some exact calculations on a chain of spins $1/2$,
  Physica {\bf 36}, 377 (1967).
  
\bibitem{Mazur-68}
  P. Mazur,
  Non-ergodicity of phase functions in certain systems,
  Physica {\bf 43}, 533 (1968).

\bibitem{McCoy-70}
  E. Barouch, B. M. McCoy, and M. Dresden,
  Statistical Mechanics of the XY  Model. I,
  Phys. Rev. A {\bf 2}, 1075 (1970).

\bibitem{McCoy-71}
  E. Barouch, B. M. McCoy, and M. Dresden,
  Statistical Mechanics of the XY  Model. II. Spin-Correlation Functions,
  Phys. Rev. A {\bf 3}, 786 (1971).

\bibitem{Bloch-08} I. Bloch, Quantum coherence and entanglement with
  ultracold atoms in optical lattices, Nature {\bf 453}, 1016 (2008).

\bibitem{GAN-14}
  I. M. Georgescu, S. Ashhab, and F. Nori, Quantum simulation, Rev. Mod.
  Phys. {\bf 86}, 153 (2014).

\bibitem{Greiner-02} M. Greiner, O. Mandel, T. Esslinger,
  T. W. H\"ansch, and I. Bloch, Quantum phase transition from a
  superfluid to a Mott insulator in a gas of ultracold atoms, Nature
  {\bf 415}, 39 (2002).

\bibitem{Weiss-06} T. Kinoshita, T. Wenger, and D. S. Weiss, A quantum
  Newton's cradle, Nature {\bf 440}, 900 (2006).

\bibitem{Schmiedmayer-07} S. Hofferberth, I. Lesanovsky, B. Fischer,
  T. Schumm, and J. Schmiedmayer, Non-equilibrium coherence dynamics
  in one-dimensional Bose gases, Nature {\bf 449}, 324 (2007).

\bibitem{Trotzky-12} S. Trotzky, Y.-A. Chen, A. Flesch,
  I. P. McCulloch, U. Schollw\"ock, J. Eisert, and I. Bloch, Probing
  the relaxation towards equilibrium in an isolated strongly
  correlated one-dimensional Bose gas, Nat. Phys. {\bf 8}, 325 (2012).

\bibitem{Cheneau-12} M. Cheneau, P. Barmettler, D. Poletti, M. Endres,
  P. Scbau\ss, T. Fukuhara, C. Gross, I. Bloch, C. Kollath, and
  S. Kuhr, Light-cone-like spreading of correlations in a quantum
  many-body system, Nature {\bf 481}, 484 (2012).

\bibitem{Schmiedmayer-12} M. Gring, M. Kuhnert, T. Langen,
  T. Kitagawa, B. Rauer, M. Schreitl, I. Mazets, D. Adu Smith,
  E. Demler, and J. Schmiedmayer, Relaxation and Prethermalization in
  an Isolated Quantum System, Science {\bf 337}, 1318 (2012).
  
\bibitem{PSSV-11} A. Polkovnikov, K. Sengupta, A. Silva, and
  M. Vengalattore, Colloquium: Nonequilibrium dynamics of closed
  interacting quantum systems, Rev. Mod. Phys. {\bf 83}, 863 (2011).

\bibitem{NH-15}
  R. Nandkishore and D. A. Huse,
  Many body localization and thermalization in quantum statistical mechanics,
  Annu. Rev. Condens. Matter Phys. {\bf 6}, 15 (2015).

\bibitem{CTGM-15}
  A. Chiocchetta, M. Tavora, A. Gambassi, and A. Mitra,
  Short-time universal scaling and light-cone dynamics after a quench
  in an isolated quantum system in $d$ spatial dimensions,
  Phys. Rev. B {\bf 94}, 134311 (2016).

\bibitem{CC-16}
  P. Calabrese and J. Cardy,
  Quantum quenches in $1+1$ dimensional conformal field theories,
  J. Stat. Mech. (2016) 064003.

\bibitem{Heyl-17} M. Heyl, Dynamical quantum phase transitions: a
  review, arXiv:1709.07461 (2017).
  
\bibitem{CPV-14}
  M. Campostrini, A. Pelissetto, and E. Vicari,
  Finite-size scaling at quantum transitions,
  Phys. Rev. B {\bf 89}, 094516 (2014).

\bibitem{Sachdev-book} S. Sachdev, {\em Quantum Phase Transitions}, 
  (Cambridge Univ. Press, 1999).

\bibitem{CV-10} M. Campostrini and E. Vicari, Equilibrium and
  off-equilibrium trap-size scaling in one-dimensional ultracold
  bosonic gases, Phys. Rev. A {\bf 82}, 063636 (2010).

\bibitem{CCCPV-00} P. Calabrese, M. Caselle, A. Celi, A. Pelissetto,
  and E. Vicari, Nonanalyticity of the Callan-Symanzik
  $\beta$-function of two-dimensional O($N$) models, J. Phys. A {\bf
    33}, 8155 (2000); M. Caselle, M. Hasenbusch, A. Pelissetto, and
  E. Vicari, Irrelevant operators in the two-dimensional Ising model,
  J. Phys. A {\bf 35}, 4861 (2002).

\bibitem{PRV-18} A. Pelissetto, D. Rossini, and E. Vicari, Off-equilibrium
  dynamics driven by localized time-dependent perturbations at quantum
  phase transitions, Phys. Rev. B {\bf 97}, 094414 (2018).

\bibitem{RSMS-09}
  D. Rossini, S. Suzuki, G. Mussardo, and G. E. Santoro, 
  Effective thermal dynamics following a quantum quench in a spin chain,
  Phys. Rev. Lett. {\bf 102}, 127204 (2009).

\bibitem{CEF-11} P. Calabrese, F.H.L. Essler, and M. Fagotti, Quantum
  Quench in the Transverse Field Ising Chain, Phys. Rev. Lett. {\bf
    106}, 227203 (2011).

\bibitem{HPK-13} M. Heyl, A. Polkovnikov, and S. Kehrein, Dynamical
  Quantum Phase Transitions in the Transverse-Field Ising Model,
  Phys. Rev. Lett. {\bf 110}, 135704 (2013).

\bibitem{HLW-94} C. Holzhey, F. Larsen, and F. Wilczek, Geometric and
  renormalized entropy in conformal field theory, Nucl. Phys. B {\bf
    424}, 443 (1994).

\bibitem{CC-04} P. Calabrese and J. Cardy, Entanglement entropy and
  quantum field theory, J. Stat. Mech. (2004) P06002.

\bibitem{JK-04} B.-Q. Jin and V. E. Korepin, 
  Quantum Spin Chain, Toeplitz Determinants and the Fisher-Hartwig Conjecture,
  J. Stat. Phys. {\bf 116}, 79 (2004).

\bibitem{CCP-10} P. Calabrese, J. Cardy, and I. Peschel, Corrections
  to scaling for block entanglement in massive spin chains,
  J. Stat. Mech. (2010) P09003.

\bibitem{Pfeuty-70}
  P. Pfeuty, The one-dimensional Ising model with a transverse field,
  Ann. Phys. {\bf 57}, 79 (1970).

\bibitem{CJ-87} G. G. Cabrera and R. Jullien, Universality of
  Finite-Size Scaling: Role of the Boundary Conditions,
  Phys. Rev. Lett. {\bf 57}, 393 (1986); Role of the boundary
  conditions in the finite-size Ising model, Phys. Rev. B {\bf 35},
  7062 (1987).
  
\bibitem{CNPV-14} M. Campostrini, J. Nespolo, A. Pelissetto, and
  E. Vicari, Finite-size scaling at first-order quantum transitions,
  Phys. Rev. Lett. {\bf 113}, 070402 (2014).

\bibitem{Greiner_2011}
  J. Simon, W. S. Bakr, R. Ma, M. E. Tai, P. M. Preiss, and M. Greiner,
  Quantum simulation of antiferromagnetic spin chains in an optical lattice,
  Nature {\bf 472}, 307 (2011).

\bibitem{Monroe_2010}
  K. Kim, M.-S. Chang, S. Korenblit, R. Islam, E. E. Edwards, J. K. Freericks, G.-D. Lin, L.-M. Duan, and C. Monroe,
  Quantum simulation of frustrated Ising spins with trapped ions,
  Nature {\bf 465}, 590 (2010).

\bibitem{MonroeB_2010}
  E. E. Edwards, S. Korenblit, K. Kim, R. Islam, M.-S. Chang, J. K. Freericks, G.-D. Lin, L.-M. Duan, and C. Monroe,
  Quantum simulation and phase diagram of the transverse-field Ising model with three atomic spins,
  Phys. Rev. B {\bf 82}, 060412(R) (2010).

\bibitem{Monroe_2014}
  P. Richerme, Z.-X. Gong, A. Lee, C. Senko, J. Smith, M. Foss-Feig, S. Michalakis, A. V. Gorshkov, and C. Monroe,
  Non-local propagation of correlations in long-range interacting quantum systems,
  Nature {\bf 511}, 198 (2014).

\bibitem{Roos_2014}
  P. Jurcevic, B. P. Lanyon, P. Hauke, C. Hempel, P. Zoller, R. Blatt, and C. F. Roos,
  Observation of entanglement propagation in a quantum many-body system,
  Nature {\bf 511}, 202 (2014).
  
\bibitem{Rydberg_2016}
  H. Labuhn, D. Barredo, S. Ravets, S. de Leseleuc, T. Macri, T. Lahaye, and A. Browaeys,
  Tunable two-dimensional arrays of single Rydberg atoms for realizing quantum Ising models,
  Nature {\bf 534}, 667 (2016).

\end{thebibliography}
\end{document}